\documentclass{elsart}
\usepackage{amsmath}
\usepackage{amssymb}
\usepackage{graphicx}
\journal{Physics Letters A}

\begin{document}
\begin{frontmatter}

\title{New classes of exact solutions for magnetic reconnective annihilation \thanksref{talk}} 
\thanks[talk]{Article presented at the 8th Plasma Easter Meeting, Turin, 23-25 April 2003.}
\author[Ruhruni]{E. Tassi\corauthref{cor}}\ead{tassi@tp4.ruhr-uni-bochum.de}, \author{V.S. Titov} and \author {G. Hornig}
\corauth[cor]{Corresponding author.}

\address{Theoretische Physik IV, Ruhr-Universit\"at Bochum, 44780 Bochum, Germany}
\address[Ruhruni]{Theoretische Physik IV, Ruhr-Universit\"at Bochum, 44780 Bochum, Germany, Telephone number: + 49 234 3223458, Fax number: + 49 234 32 14177}  

\begin{abstract}
Analytical solutions for reconnective annihilation in curvilinear geometry are presented. These solutions are characterized by current density distributions varying both along the radial and azimuthal coordinates. They represent an extension of previously proposed models, based on purely radially dependent current densities. Possible applications of these solutions to the modeling of solar flares are also discussed.
\end{abstract}

\end{frontmatter}

\section{Introduction}

Magnetic reconnection is a process in which magnetic energy is converted into kinetic and thermal energy of a plasma. Its importance lies in the fact that magnetic reconnection is believed to be one of the underlying mechanisms responsible for many phenomena occurring in astrophysical plasmas, such as solar flares and geomagnetic substorms \cite{PrFbs00,Bi00}. After the early works of Parker \cite{Pa57}, Sweet \cite{Sw58} and Petschek \cite{Pe64} considerable effort has been made to improve the theory of magnetic reconnection with the help of exact analytical models for reconnective annihilation in Cartesian coordinates \cite{So75,Cr95,Pr00}. More recently exact solutions for magnetic reconnection in curvilinear coordinates have been presented \cite{Wa02,Ta02,Ta03}. Some features of these solutions make them particularly interesting for modeling a large class of solar flares. In this paper we present analytical two-dimensional solutions for reconnective annihilation in polar coordinates which extend the model derived in \cite{Wa02,Ta02} to the case of current density distributions depending on both the radial and the azimuthal variable.

\section{Basic equations}
We consider the system of magnetohydrodynamics equations in polar coordinates for a two-dimensional steady and incompressible plasma with uniform density and resistivity discussed in \cite{Ta02}. This system consists of the equation of motion
\begin{equation}    \label{e:mot2}
[\psi,{\nabla}^2 \psi]=[A,{\nabla}^2 A],
\end{equation}
and the Ohm's law
\begin{equation}    \label{e:Ohm2}
E+\frac{1}{r}[\psi,A]=-\eta{\nabla}^2 A.
\end{equation} 
The flux function $A$ and the stream function $\psi$ are related to the radial and azimuthal components of the magnetic and velocity field, respectively, in the following way:
\begin{equation}
(v_{\mathrm{r}}, v_{\theta})=\left(\frac{1}{r}{\frac{\partial \psi}{\partial \theta}}, -{\frac{\partial \psi}{\partial r}}\right), \qquad (B_r, B_{\theta})=\left(\frac{1}{r}{\frac{\partial A}{\partial \theta}}, -{\frac{\partial A}{\partial r}}\right).
\end{equation}
In (\ref{e:Ohm2}) $E$ represents the normalized electric field which is uniform and perpendicular to the $r$-$\theta$ plane while $\eta$ is the dimensionless resistivity corresponding to the inverse magnetic Reynolds number. The Poisson brackets are here defined as
\begin{equation}
[f,g]={\frac{\partial f}{\partial r}}{\frac{\partial g}{\partial \theta}}-{\frac{\partial g}{\partial r}}{\frac{\partial f}{\partial \theta}}.
\end{equation}
As shown in \cite{Wa02,Ta02} the ansatz
\begin{equation}  \label{e:ansA}
A(r,\theta)=A_1(r)\theta+A_0(r),
\end{equation}
\begin{equation}  \label{e:ansP}
\psi(r,\theta)=\psi_1(r)\theta+\psi_0(r),
\end{equation}
is compatible with the system of equations (\ref{e:mot2}) and (\ref{e:Ohm2}). In fact the substitution of (\ref{e:ansA}) and (\ref{e:ansP}) into (\ref{e:mot2}) and (\ref{e:Ohm2}) yields the following set of four ordinary differential equations for four unknown functions:
\begin{equation}  \label{e:i1}
\frac{{\psi_1}^{'}}{r}{\left(r{\psi_1}^{'}\right)}^{'}
-\psi_1{\left[{\frac{1}{r}}{(r{\psi_1}')}^{'}\right]}^{'}=
{\frac{{A_1}'}{r}}{\left({r{A_1}'}\right)}^{'}-A_1{\left[{\frac{1}{r}}{(r{A_1}')}^{'}\right]}^{'},
\end{equation}
\begin{equation}  \label{e:i2}
{\frac{{\psi_0}'}{r}}{\left({r{\psi_1}'}\right)}'-\psi_1{\left[\frac{1}{r}{(r{\psi_0}')}'\right]}^{'}={\frac{{A_0}'}{r}}{\left({r{A_1}'}\right)}^{'}-A_1{\left[\frac{1}{r}{(r{A_0}')}^{'}\right]}^{'},
\end{equation}
\begin{equation}  \label{e:i3}
  \psi_1 'A_1-\psi_1 A_1 '+\eta ({A_1}^{\prime}+ r {A_1}^{\prime\prime})=0,
\end{equation}
\begin{equation}  \label{e:i4}
E+\frac{1}{r}[\psi_0 'A_1-\psi_1 A_0 '+\eta ({A_0}^{\prime}+ r {A_0}^{\prime\prime})]=0.
\end{equation}
where the dash indicates the derivative with respect to $r$. In the limit of ideal MHD, corresponding to $\eta=0$, equation (\ref{e:i3}) implies 
\begin{equation} \label{e:prop}
\psi_1=\alpha A_1
\end{equation}
with $\alpha$ being an arbitrary constant. Combining this proportionality relation with Eq. (\ref{e:i1}) yields
\begin{equation}  \label{e:eqA1}
{A_1}^{\prime\prime}+\frac{{A_1}^{\prime}}{r}\pm{\lambda}^2 A_1=0,
\end{equation}
where $\pm{\lambda}^2=-\left.\left({B_{\mathrm{r}}}^{-1} \partial j/ \partial \theta \right)\right\vert_{r=1}$ is a parameter related to the variation of the current density with respect to the variable $\theta$. For $\lambda=0$ the current density depends only on $r$ and the solution of (\ref{e:eqA1}) is given by
\begin{equation}   \label{e:br0}
A_1=c_1 \ln r +c_2.
\end{equation}
For this particular case approximate and exact solutions of the system (\ref{e:i1})-(\ref{e:i4}) have been derived (see \cite{Ta02} and \cite{Wa02} respectively). For $\lambda \neq 0$ the solutions of (\ref{e:eqA1}) depend on the sign of $\pm{\lambda}^2$ and they are given by 
\begin{equation}   \label{e:br+}
A_1^+=c^+_1 J_0(\lambda r)+c^+_2 Y_0(\lambda r),
\end{equation}
\begin{equation}   \label{e:br-}
A_1^-=c^-_1 I_0(\lambda r)+c^-_2 K_0(\lambda r).
\end{equation}
where the superscripts of the arbitrary constant $c_1$ and $c_2$ correspond to the sign of $\pm{\lambda}^2$. These solutions have an analogous counterpart in the trigonometric and hyperbolic sinus solutions derived in \cite{Pr00} for the corresponding problem in Cartesian system of coordinates. The ideal solutions for ${A_0}^{\prime}$ and ${\psi_0}^{\prime}$ can be derived from (\ref{e:i2}) and (\ref{e:i4}) setting $\eta=0$ which yields
\begin{equation}  \label{e:idA0f0}
{\psi_0}^{\prime}=\frac{1}{\alpha} {A_0}^{\prime}+\frac{\alpha^2-1}{\alpha^2}\left(a r+\frac{b}{r}\right), \qquad {A_0}^{\prime}=\frac{\alpha}{{\alpha}^2-1}{\frac{Er}{A_1}}+\frac{a}{\alpha}r+\frac{b}{{\alpha r}},
\end{equation}
with $a$ and $b$ arbitrary constants. These solutions are singular in correspondence to the zeros of $A_1$, therefore they have an infinite number of singularities if the solution $A_1^+$ is considered and only one singularity if $\pm {\lambda}^2 \leqslant 0$. It should be mentioned that in absence of an electric field the system possesses regular solutions describing field-aligned flows with one or an infinite number of null and stagnation points depending on the sign of $\pm{\lambda}^2$. Singularities in the ideal solutions (\ref{e:idA0f0}) can be resolved by means of a finite resistivity. Here we derive solutions for $\lambda \neq 0$, characterized by current density distributions depending on both $r$ and $\theta$, unlike the case $\lambda=0$ where the current density is purely radially dependent.

\section{Solutions for $A_1$ and $\psi_1$}  \label{sec:A1f1}

For the functions $A_1$ and $\psi_1$ we assume the following boundary conditions:
\begin{equation} \label{e:obc}
A_1(1)=B_{\mathrm{re}}, \qquad \psi_1(1)=v_{\mathrm{re}},
\end{equation}
\begin{equation} \label{e:ibc1}
A_1(r_{\mathrm{c}})=0, \qquad \psi_1(r_{\mathrm{c}})=0,
\end{equation}
where $B_{\mathrm{re}}$ and $v_{\mathrm{re}}$ represent constant values of the radial component of the magnetic and velocity field imposed at the radius $r=1$, while $r_{\mathrm{c}}$ indicates the radius at which $A_1^{\pm}$ and $\psi_1^{\pm}$ have a zero. The ideal solutions (\ref{e:br+}) and (\ref{e:br-}), as in the case of the logarithmic solution (\ref{e:br0}), can satisfy all of the four boundary conditions (\ref{e:obc}) and (\ref{e:ibc1}) taking the form 
\begin{equation} \label{e:A1pbc}
A_1^+=B_{\mathrm{re}}\frac{J_0(\lambda r_{\mathrm c})Y_0(\lambda r)-Y_0(\lambda r_{\mathrm c})J_0(\lambda r)}{Y_0(\lambda)J_0(\lambda r_{\mathrm c})-Y_0(\lambda r_{\mathrm c})J_0(\lambda)}
\end{equation}
and
\begin{equation}   \label{e:A1mbc}
A_1^-=B_{\mathrm{re}}\frac{I_0(\lambda r_{\mathrm c})K_0(\lambda r)-K_0(\lambda r_{\mathrm c})I_0(\lambda r)}{K_0(\lambda)I_0(\lambda r_{\mathrm c})-K_0(\lambda r_{\mathrm c})I_0(\lambda)},
\end{equation}
respectively, together with 
\begin{equation}  \label{e:f1pmbc}
{\psi_1}^{\pm}=\frac{v_{\mathrm re}}{B_{\mathrm re}}A_1^{\pm}. 
\end{equation}
A fundamental difference between the cases $\lambda=0$ and $\lambda \neq 0$ is that for the former the ideal solutions for $A_1$ and $\psi_1$ provided by (\ref{e:prop}) and (\ref{e:br0}) are also exact solutions of the complete resistive equations (\ref{e:i1}) and (\ref{e:i4}). For $\lambda \neq 0$, however, this is not the case. In fact, comparing the Taylor expansion about $r_{\mathrm c}$ of the exact solutions of (\ref{e:i1}) and (\ref{e:i3}) with the one of the ideal solutions it transpires that a mismatch occurs between $A_1^{\pm \prime\prime\prime}(r_{\mathrm c})$ and the third derivative of the exact solution for $A_1$ at $r_{\mathrm c}$ . Due to the high complexity of the nonlinear equations (\ref{e:i1}) and (\ref{e:i3}) we limit ourselves to solve them in an approximated way. In particular we seek solutions satisfying the imposed boundary conditions and whose Taylor expansions about $r_{\mathrm c}$ are compatible with the ones of the exact solutions at least up to the third derivative, in order to resolve the lowest order mismatch between ideal and exact solutions. Since the typical value of the dimensionless resistivity $\eta$ for many astrophysical and laboratory plasmas is between $10^{-10}$ and $10^{-14}$ we can consider $\eta$ as a small parameter and use the method of matched asymptotic expansions. We separate the problem in two domains, an inner region consisting of a narrow layer containing $r_{\mathrm c}$, where the ideal solution breaks down, and an outer region, complementary to the inner region, where (\ref{e:A1pbc}) and (\ref{e:A1mbc}) are approximate solutions. Asymptotic expansions for $A_1$ and $\psi_1$ are found in the inner region and  then matched with the outer zeroth order expansion (\ref{e:A1pbc}), (\ref{e:A1mbc}) and (\ref{e:f1pmbc}). In order to derive inner solutions of the nonlinear equations (\ref{e:i1}) and (\ref{e:i3}) the dependent and independent variables are rescaled in the following way: 
\begin{equation}
A_1=\sqrt{\eta}r_{\mathrm c}\bar{A_1}, \qquad  \psi_1=\sqrt{\eta}r_{\mathrm c}\bar{\psi_1}, \qquad
s=\frac{r-r_{\mathrm c}}{\sqrt{\eta}}.
\end{equation}
The equations obtained after the rescaling are
\begin{equation} \label{e:i1rsc}
\begin{split}
&\left[{\left({\frac{\bar{\psi_1}}{\bar{\psi_1}^{\prime\prime}}}\right)}^{\prime}{{\bar{\psi_1}}}^{\prime\prime2}-{\left({\frac{\bar{A_1}}{\bar{A_1}^{\prime\prime}}}\right)}^{\prime}{\bar{A_1}}^{\prime\prime2}\right]+\\
&\frac{\sqrt{\eta}}{r_{\mathrm c}}\left[\bar{\psi_1}^{\prime 2}{\left({\frac{\bar{\psi_1}}{\bar{\psi_1}^{\prime}}}\right)}^{\prime}+2s{{\bar{\psi_1}}}^{\prime\prime2}{\left({\frac{\bar{\psi_1}}{\bar{\psi_1}^{\prime}}}\right)}^{\prime}-\bar{A_1}^{\prime 2}{\left({\frac{\bar{A_1}}{\bar{A_1}^{\prime}}}\right)}^{\prime}-2s{{\bar{A_1}}}^{\prime\prime2}{\left({\frac{\bar{A_1}}{\bar{A_1}^{\prime}}}\right)}^{\prime}\right]+\\
&\frac{\eta}{{r_{\mathrm c}}^2}\left[\bar{\psi_1}\bar{\psi_1}^{\prime}+s\bar{\psi_1}^{\prime 2}{\left({\frac{\bar{\psi_1}}{\bar{\psi_1}^{\prime}}}\right)}^{\prime}+s^2{{\bar{\psi_1}}}^{\prime\prime2}{\left({\frac{\bar{\psi_1}}{\bar{\psi_1}^{\prime\prime}}}\right)}^{\prime}-\bar{A_1}\bar{A_1}^{\prime}-s\bar{A_1}^{\prime 2}{\left({\frac{\bar{A_1}}{\bar{A_1}^{\prime}}}\right)}^{\prime} \right.\\
& \left. -s^2{{\bar{A_1}}}^{\prime\prime2}{\left({\frac{\bar{A_1}}{\bar{A_1}^{\prime\prime}}}\right)}^{\prime}\right],\\
\end{split}
\end{equation}
\begin{equation} \label{e:i2rsc}
{\bar{\psi_1}}^{\prime}\bar{A_1}-{\bar{A_1}}^{\prime}\bar{\psi_1}+\bar{A_1}^{\prime\prime}+\frac{\sqrt{\eta}}{r_{\mathrm c}}(s\bar{A_1}^{\prime\prime}+{\bar{A_1}}^{\prime}).
\end{equation}
It should be noted that in the limit $r_{\mathrm c} \rightarrow \infty$ the above equations reduce to the rescaled equations for the problem in Cartesian coordinates considered in \cite{Pr00}. An asymptotic expansion in powers of $\eta$ is chosen as a form of solutions for (\ref{e:i1rsc}) and (\ref{e:i2rsc}). Also in curvilinear geometry the rescaled equations depend explicitly on $\eta$ and in particular the square root of $\eta$ is present in them. Therefore it is natural to choose inner solutions of the following form:
\begin{equation} 
\begin{split}
&\bar{A_1}=\bar{A_{10}}(s)+\sqrt{\eta}\bar{A_{11}}(s)+\eta\bar{A_{12}}(s)+\cdots\\
&\bar{\psi_1}=\bar{\psi_{10}}(s)+\sqrt{\eta}\bar{\psi_{11}}(s)+\eta\bar{\psi_{12}}(s)+\cdots\\
\end{split}
\end{equation}
where, unlike to the case of Cartesian geometry, the square root of $\eta$ appears in the expansions. The resulting inner solutions are then written in terms of the original variable, expanded as $\eta \rightarrow 0$ for fixed $r$ and matched with the expansion about $r_{\mathrm c}$ of the outer solutions (\ref{e:A1pbc}) and (\ref{e:A1mbc}). Eventually one obtains 
\begin{equation}  \label{e:A1f1appr}
\begin{split}
& \bar{A_1}^{\pm}=\frac{B_{\mathrm{re}}\sigma^{\pm}}{r_{\mathrm c}}s-\frac{\sqrt{\eta}}{r_{\mathrm c}}\frac{B_{\mathrm{re}}\sigma^{\pm}}{2 r_{\mathrm c}}s^2+\eta \left\{\frac{B_{\mathrm{re}}\sigma^{\pm} C^{\pm}}{6r_{\mathrm c} k^{\pm}}s^3+s\left(\frac{2 B_{\mathrm{re}}\sigma^{\pm}}{k^{\pm} r_{\mathrm c}^3}-\frac{B_{\mathrm{re}}\sigma^{\pm} C^{\pm}}{r_{\mathrm c} {k^{\pm}}^2}\right)\right.\\
&\left. \left[\sqrt{\frac{2}{k^{\pm}}}\frac{1}{s}\rm{daw}\left(\sqrt{\frac{k^{\pm}}{2}}s\right)-1+\sqrt{2}\int_0^{\sqrt{k^{\pm}}s}{\rm{daw}\left(\frac{t}{\sqrt{2}}\right)dt}\right]\right\},\\
& \bar{\psi_1}^{\pm}=\frac{v_{\mathrm{re}}\sigma^{\pm}}{r_{\mathrm c}}s-\frac{\sqrt{\eta}}{r_{\mathrm c}}\frac{v_{\mathrm{re}}\sigma^{\pm}}{2 r_{\mathrm c}}s^2+\eta \left\{\frac{v_{\mathrm{re}}\sigma^{\pm} C^{\pm}}{6r_{\mathrm c} k^{\pm}}s^3+s\left(\frac{2 {B_{\mathrm{re}}}^2\sigma^{\pm}}{v_{\mathrm{re}}k^{\pm} r_{\mathrm c}^3}-\frac{{B_{\mathrm{re}}}^2\sigma^{\pm} C^{\pm}}{v_{\mathrm{re}}r_{\mathrm c} {k^{\pm}}^2}\right)\right.\\
 &\left. \left[\sqrt{\frac{2}{k^{\pm}}}\frac{1}{s}\rm{daw}\left(\sqrt{\frac{k^{\pm}}{2}}s\right)-1+\sqrt{2}\int_0^{\sqrt{k^{\pm}}s}{\rm{daw}\left(\frac{t}{\sqrt{2}}\right)dt}\right]\right\},\\
\end{split}
\end{equation}
with
\begin{equation}
\begin{split}
&\sigma^{+}=\lambda \frac{Y_0(\lambda r_{\mathrm c})J_1(\lambda r_{\mathrm c})-J_0(\lambda r_{\mathrm c})Y_1(\lambda r_{\mathrm c})}{Y_0(\lambda)J_0(\lambda r_{\mathrm c})-Y_0(\lambda r_{\mathrm c})J_0(\lambda)},\\ 
&\sigma^{-}=\lambda \frac{K_0(\lambda r_{\mathrm c})I_1(\lambda r_{\mathrm c})+I_0(\lambda r_{\mathrm c})K_1(\lambda r_{\mathrm c})}{I_0(\lambda)K_0(\lambda r_{\mathrm c})-I_0(\lambda r_{\mathrm c})K_0(\lambda)},\\
&C^{\pm}=\frac{2 a_1^2}{r_{\mathrm c}^3 p_1}-\frac{p_3}{r_{\mathrm c}}k^{\pm}.\\
\end{split}
\end{equation}
and where 
\begin{equation}
k^{\pm}=\frac{{B_{\mathrm{re}}}^2-{v_{\mathrm{re}}}^2}{v_{\mathrm{re}}r_{\mathrm c}}\sigma^{\pm}
\end{equation}
is a positive quantity. The inner solutions (\ref{e:A1f1appr}) possess the required properties, since they satisfy the boundary conditions (\ref{e:ibc1}) and  $\bar{A_1}^{\pm \prime\prime\prime}(r_{\mathrm c})$ is consistent with the value of the third derivative of the exact solution at $r_{\mathrm c}$. By combining (\ref{e:A1f1appr}) with (\ref{e:A1pbc}), (\ref{e:A1mbc}) and (\ref{e:f1pmbc}) composite solutions can be found, which are approximately valid over the whole domain of interest.

\section{Solutions for ${A_0}^{\prime}$ and ${\psi_0}^{\prime}$ } \label{sec:A0f0}

As boundary condition for ${A_0}^{\prime}$ we impose
\begin{equation} \label{e:ibc0}
{A_0}^{\prime}(r_{\mathrm{c}})=0.
\end{equation}
This constraint, together with (\ref{e:ibc1}), fixes the position of a magnetic null at the point $(r=r_{\mathrm c}, \theta=0)$. On the other hand, as mentioned in Sec. \ref{sec:A1f1} the ideal solutions (\ref{e:idA0f0}) become singular at $r_{\mathrm c}$ which is in striking contrast with the condition (\ref{e:ibc0}). Therefore the ideal solution for ${A_0}^{\prime}$ and ${\psi_0}^{\prime}$ also breaks down at $r=r_{\mathrm c}$ and a boundary layer analysis is required in order to find approximate resistive inner solutions satisfying (\ref{e:ibc0}). The inner equations for ${A_0}^{\prime}$ and ${\psi_0}^{\prime}$ are derived by rescaling the independent variable in Eqs. (\ref{e:i2}) and (\ref{e:i4}) in the following way:
\begin{displaymath}
s=\frac{r-r_{\mathrm{c}}}{\sqrt{2 \eta}}.
\end{displaymath}
Then an asymptotic expansion in powers of $\eta$ is sought in the form
\begin{equation} 
{A_0}^{\prime}={A^{0}_0}^{\prime}+\sqrt{\eta}{A^{1}_0}^{\prime}, \qquad {\psi_0}^{\prime}={\psi^{0}_0}^{\prime}+\sqrt{\eta}{\psi^{1}_0}^{\prime}.
\end{equation}
In the equations for the inner solutions $A_1$ and $\psi_1$ are replaced by the leading terms of the Taylor expansions of (\ref{e:A1f1appr}) about $s=0$. The resulting equations are formally equivalent to the corresponding equations solved in \cite{Ta02}. The composite solutions obtained after the matching with (\ref{e:idA0f0}) are given by
\begin{eqnarray}
&{A_0}^{\prime} = E \left[ {\rm{daw}} (\sqrt{k^{\pm}}s) \left(-\frac{2}{3}\frac{\sqrt{k^{\pm}}s^3}{r_{\mathrm{c}}} +\frac{\sqrt{2}}{r_{\mathrm{c}} \sqrt{k^{\pm}}}s^2+  \frac{2}{\sqrt{k^{\pm}}r_{\mathrm{c}}}s+\frac{1}{2r_{\mathrm{c}} k^{\pm}}-\frac{1}{\sqrt{\eta}}\sqrt{\frac{2}{k^{\pm}}}\right) \right.  \nonumber \\
& \left. +\frac{1}{{r_{\mathrm{c}}k^{\pm}}}\left(\frac{7}{3}\mathrm{e}^{-k^{\pm}s^2}+\frac{k^{\pm}}{3}s^2 -\frac{s}{\sqrt{2}}-\frac{5}{6}\right)+\frac{1}{k^{\pm}\sqrt{2\eta}s}+\frac{s\sqrt{2\eta}+r_{\mathrm{c}}}{k^{\pm}{r_{\mathrm{c}}}^2\ln\left(\frac{r_{\mathrm{c}}}{s\sqrt{2\eta}+r_{\mathrm{c}}}\right)}\right] \nonumber \\
& -\left(a{r_{\mathrm{c}}}+\frac{b}{{r_{\mathrm{c}}}}\right)\frac{B_{\mathrm{re}}}{v_{\mathrm{re}}}\mathrm{e}^{-k^{\pm}s^2}+\left[a(s\sqrt{2\eta}+{r_{\mathrm{c}}})+\frac{b}{s\sqrt{2\eta}+r_{\mathrm{c}}}\right]\frac{B_{\mathrm{re}}}{v_{\mathrm{re}}} \nonumber \,
\end{eqnarray}
\begin{equation}
{\psi_0}^{\prime}=\frac{B_{\mathrm{re}}}{v_{\mathrm{re}}}{A_0}^{\prime}+\left[1-{\left(\frac{B_{\mathrm{re}}}{v_{\mathrm{re}}}\right)}^2\right]\left( a{\it r_{\mathrm{c}}}+{\frac {b}{{\it r_{\mathrm{c}}}}} \right). \nonumber
\end{equation}

\section{Discussion of the solutions}

The solutions derived in Sec. \ref{sec:A1f1} and \ref{sec:A0f0} describe stationary magnetic and velocity fields whose topologies are nontrivial due to the presence of a magnetic null point at $(r=r_{\mathrm{c}}, \theta= 0)$ and of a stagnation point at $(r=r_{\mathrm{c}}, \theta=[(-{{v_{\mathrm{re}}}^2-{B_{\mathrm{re}}}^2})/{{v_{\mathrm{re}}}^3 \sigma^{\pm}}](a r_{\mathrm{c}}+{b}/{r_{\mathrm{c}}})$. Each of these singular points is located at the intersection of two separatrix lines, one of which, corresponding to the arc $r=r_c$, is in common to both points. Along this shared separatrice a curved sheet of enhanced current density is present, allowing for reconnection of magnetic field lines that are advected by the flow from the outer region. The rate at which reconnection occurs is given by the parameter $E$ that does not depend on $\eta$. Thus these solutions belong to the class of fast reconnection models. As in the case of many other magnetohydrodynamics models of reconnective annihilation \cite{Cr95,Pr00,Wa02,Ja92} a limit to the reconnection rate is imposed by the requirement of incompressibility. Indeed, consider the expression for the plasma pressure given by
\begin{equation}   \label{e:press}
p=-\frac{{v_r}^2}{2}-\frac{{B_{\theta}}^2}{2}-\int{\frac{{B_{\theta}}^2-{v_{\theta}}^2}{r}dr}+p_0
\end{equation}
with constant $p_0$. The second term on the right hand side of (\ref{e:press}) is negative and diverges as $r \rightarrow \infty$. Therefore, in order for $p$ to be a positive quantity, the domain of the solutions must be limited to a finite region of space.\\
The current density is perpendicular to the $r$-$\theta$ plane and its expression is given by 
\begin{equation}
j=-\left(\frac{A_1^{\prime}}{r}+A_1^{\prime\prime}\right)\theta-\frac{{A_0}^{\prime}}{r}-{A_0}^{\prime\prime},
\end{equation}
from where it is possible to notice the linear dependence of $j$ on $\theta$. With the help of (\ref{e:eqA1}) it is also possible to see that this dependence is absent for $\lambda=0$. 

\begin{figure}  
\begin{center}
\includegraphics[width=15cm]{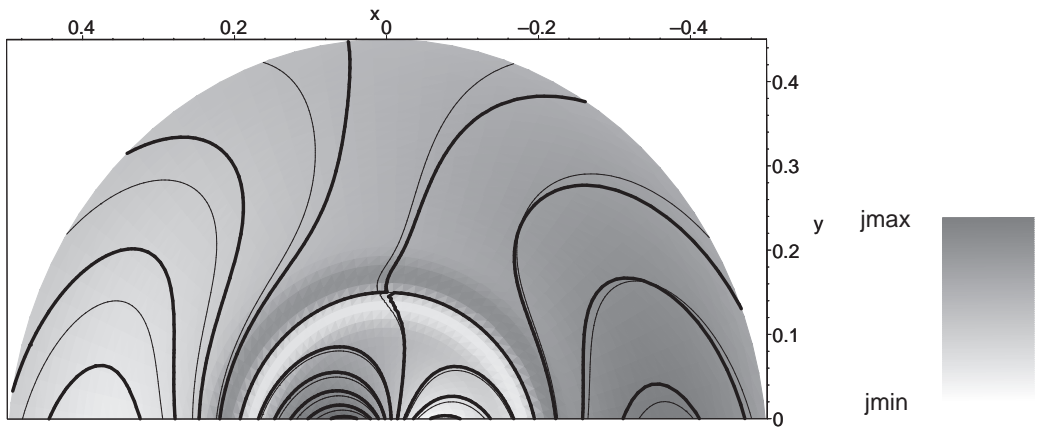}
\end{center}
\caption{Magnetic field lines (thick) and streamlines (thin) for $\lambda=8.7$, $E=0.5$, $\eta=10^{-2}$, $v_{\mathrm{re}}=0.8$, $B_{\mathrm{re}}=0.9$, $r_{\mathrm{c}}=0.2$, $a=0$ and $b=1$. The plots of the field lines are superimposed to the distribution of the current density in gray half-tones.}
\label{fig:magvel}
\end{figure}

The variation of the current density with $\theta$ can clearly be seen in Fig. \ref{fig:magvel}, which shows that this variation is considerably large in the outer region, while in the inner region the current density is practically only radially dependent.\\ 
Finally, from Fig. \ref{fig:magvel} one can also notice that for the branch of solutions matching ${A_1}^+$ and ${\psi_1}^+$ the resulting magnetic configuration can be considered as generated by an odd number of sources with alternating polarities lying on the plane $y=0$. The number of sources depends on the value of $\lambda$ and on how large the domain of validity of the solution is. The latter cannot be chosen arbitrarily large because it would then include additional zeros of (\ref{e:br+}) where the solutions become singular. However this is not an additional restriction. Indeed, as already mentioned above, the exact magnetohydrodynamic solutions for incompressible plasma are usually valid only in a finite size region anyway because of the lower limitation on the pressure. On the other hand, the magnetic and flow configurations corresponding to $\pm \lambda^2 <0$ resemble very much the ones for $\lambda=0$ solutions, which are regular everywhere and therefore they are not subject to restrictions due to the presence of singularities. The current density for this branch of solutions varies also linearly with $\theta$ but in the opposite direction to the case $\pm \lambda^2 >0$.   

\section{Conclusions}
Solutions of the steady two dimensional MHD equations for an incompressible plasma in polar coordinates have been derived using a matched asymptotic expansion technique where the inverse magnetic Reynolds number has been assumed to be very small. Two branches of solutions have been obtained depending on the value of one parameter related to the variation of the current density with the azimuthal variable. The magnetic field configuration described by the solutions has a null point at a certain radius. Along this radius a curved current layer is formed, which causes the reconnection of field lines. In the magnetic configuration three alternating polarities lying on the plane $y=0$ produce the magnetic field with the null point in the volume above. This feature has interesting applications to solar physics because the presence of three alternating polarities on the photospheric plane is a feature common to a large class of solar flares \cite{Ni97}. For one branch of solutions, in general, on the plane $y=0$ an odd number of sources is present, three of which interacting with each other. This number depends on the value of the parameter $\lambda$ and on the size of the domain of validity of the solutions. The reconnection rate does not depend on the resistivity therefore these solutions belong to the class of models providing fast reconnection regimes.

\bibliographystyle{plain}

\end{document}